# ENRICHING INFORMATION TECHNOLOGY COURSE MATERIALS BY USING YOUTUBE

**Leon Andretti Abdillah**
Faculty of Computer Science
Universitas Bina Darma
*leon.abdillah@yahoo.com*

**Abstract**

*IT offers some benefits and collaborations in various sectors. This research focuses on exploring higher education subjects via social technology, YouTube. YouTube is the world largest video based contents application in the world. Current learning materials are not only in text and images, but included video contents. This research enriching students learning materials may involving YouTube as learning sources. The study observed 118 sophomore students in computer science faculty. The results show that, involving YouTube in enriching students course material able to create conductive learning environment. This stratey increases students' understanding in their field of study.*

**Keywords:**    *Socal Technology, Social Media, YouTube, SCM.*

---------------------------------------------------------------------------------------------------------------------

## 1. Introduction

Cloud social technology as part of information technology (IT) applications has been growed tremendously recently. The users of those social technologies have been reached millions of people. Among the diverse applications of emerging IT, group of social technology or social media has been attracked the author's attention for a long time. At the moment, there are 15 most popular social networking (SN) sites (Kallas, 2017). Facebook and YouTube lead in the top first and second place. These famous SN applications have been used in many sectors, such as business, banking, government, medicine, even to the field of education. IT has the capability to collaborate (Abdillah, Syafe'i, & Hardiyansyah, 2007) with all those fields. A number of the most popular IT applications have been so commonplace in the field of education has brought some remarkable progress.

The development of information and communication technology has been influenced the learning process nowadays in higher education or universities. Conventional teaching methods have undergone significant changes. Delivery of course materials are not enough to face to face with the help of the projector. Changing times have led demands more competition. Lecturers are required to innovate in enriching the learning method involving a number of IT applications. In IT trend, one of which is a social technology that facilitates connections between individuals in the real world. People today have been able to relate to others as if without limits of space and time.

Numerous studies have been commenting about the social technologies in various fields: 1) promotion in business (Rahadi & Abdillah, 2013), 2) political campaign and presidential election (Abdillah, 2014a, 2014c), 3) knowledge management systems (Abdillah, 2014b), 4) famous Indonesan online transportation of "Go-jek" (Oktaviana, Syah, & Abdillah, 2016; Wulandari, Syah, & Abdillah, 2016), and 5) education (Abdillah, 2013, 2016a, 2016b). The focus of current research is the utilization of social technology, Facebook and YouTube, in enriching course material for students in higher education sector.

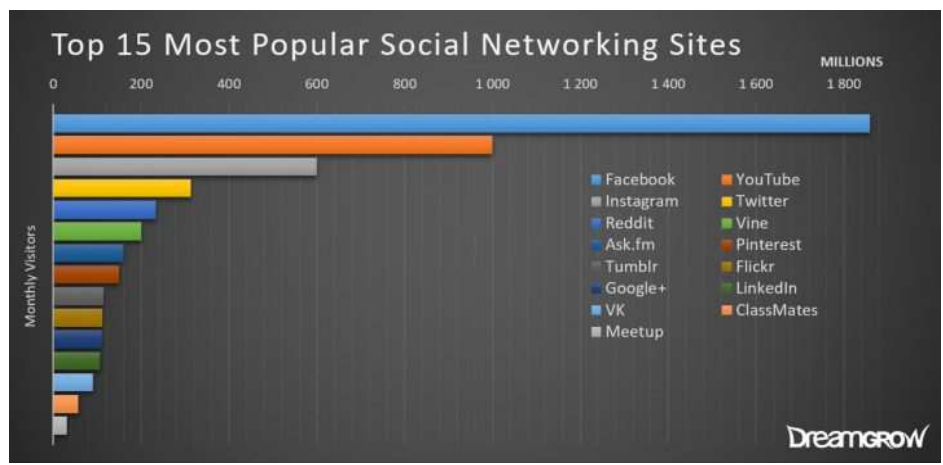

Figure 1: Top 15 Most Popular Social Netwoorking Apps.

YouTube is one of the most well-known and widely discussed sites of participatory media in the contemporary online environment, and it is the first genuinely mass-popular platform for user-created video (Burgess & Green, 2013). Emerging technologies, such as the YouTube video-sharing Web site, are important for both in-class and online instructors to establish a sense of classroom community and achieve greater learner outcomes (Burke & Snyder, 2008). YouTube is the most famous education channel (Saurabh & Sairam, 2013). The use of video in education can be an effective way of engaging students and supporting their understanding (Tan & Pearce, 2012). Student easiness in following the lecture today increasingly rely on information technology that is where one of them in obtaining teaching materials lecture (Pradikta & Haryono, 2015).

The main concern of this study is enrichment of lecture materials for "business modeling - supply chain management" courses. The next section of this article contains the research methods used for this research to work well. In the third section, contains a discussion of the aspects that exist within the scope of research, and closed with a number of conclusions obtained after the study was conducted.

## 2. Methodology

### 2.1 Sample and data collection method

The method used in this research is case study method. The object of research is the subject of "business modeling - supply chain management". This course is one of the subjects offered in the information system study program (SI) in the computer science faculty. The students are second year students (sophomore) or in semester 4 (four). These students have grown up within a world of pervasive technology including mobile phones, digital cameras and the omnipresent internet (Duffy, 2008). The total respondents involved were 112 students. All of the respondents are asked to join with Facebook group that has been created before by lecturer.

### 2.2 Procedures

Lecturer needs to browse some of YouTube channels that provide the contents in course subject of "business modeling – supply chain management". Next, every class is given with YouTube's URL of such a topic. In this research, the numbers of unified resource locators (URLs) are adjusted with the number of groups in the class. If in a class consist of 5 (five) groups, then lecturer will provides

YouTube's URL as much as group number in that class. According to current data, the number of group in every class is ranged from 4 (four) to 6 (six) groups. Then, lecturer will provided 6 (six) YouTube's URLs. For this case study, lecturers take samples on 3 (three) lecture topics. The three topics are: 1) Transportation Topics, 2) Vendor Management Topics, and 3) Procurement Topics.

Announcements on task instructions were sent to study groups on Facebook. To make it easier for students to make a summary, the lecturer prepares an empty report template containing the summary report framework. The report template will contain a number of sections according to the number of students in the group. Reports that have been done will be stored in the cloud repository, DropBox. DropBox is a free software tool that allows anyone to synchronize files across home and office computers, smartphones, and tablets ( Ries, Aliponga, & Kochiyama, 2012). Each representative of each group will then submit their group report by pasting the URL from their group DropBox.

### 3. Finding & Discussion

### 3.1 Respondents Characteristics

Total registered students in this research are 118 students. Those students enrolled in facebook group of "Business Modeling – Supply Chain Management". Among those registered students, 112 responses into the test. Male students dominated the study or equal to 87.5%. While the rest are female students totaling 12.5 percent (see Figure 2).

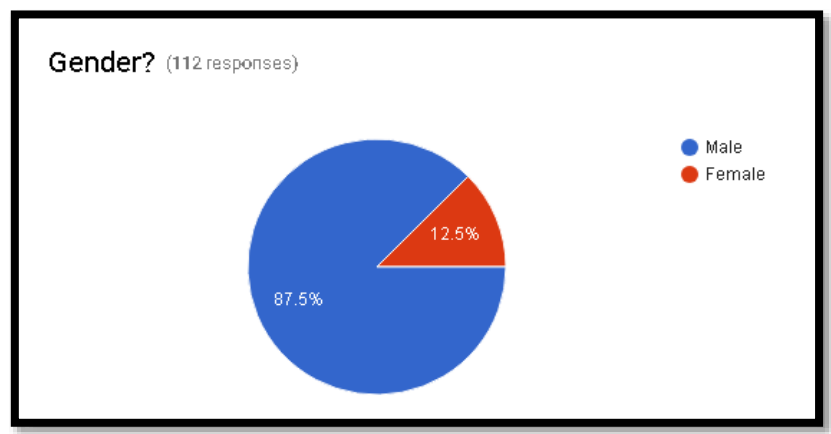

Figure 2: Respondents Characteristics (Gender).

### 3.2 Transportation

The first focus topic is "Transportation". Distribution from raw materials into finished product involved multimodal transports. This is one important part of supply chain management. The transportation of a good could be used many mode of transport. Figure 4 shows an example of the transportation chain (Pillai, 2012).

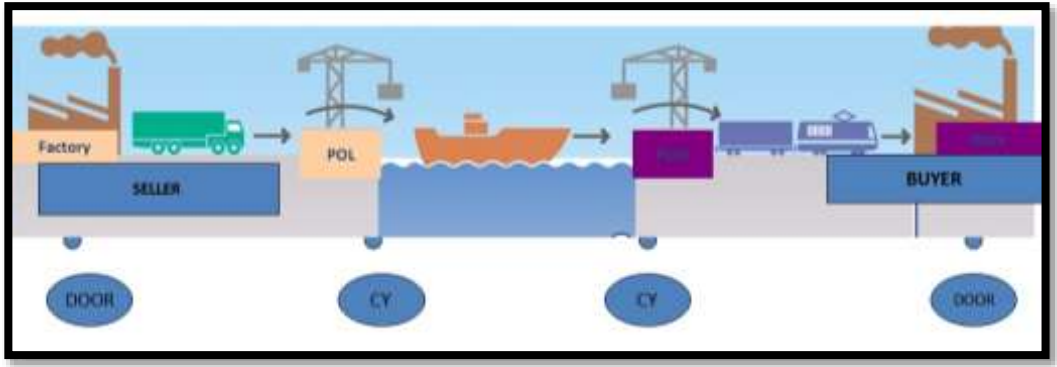

Figure 3: Transportation Chain.

Every group of students are asked to browse the topic of "Transportation in SCM". There are 6 (six) materials related to "Transportation", as follow: 1) "Move It: Transportation and Logistics" (Carey, 2010), 2) "Introduction to Supply Chain Transportation" (Harrod, 2012), 3) "Supply Chain Management : Modes of Transportation" (Edupedia World, 2015a), 4) "Total Logistic Control : Transportation & Supply Chain 3PL Management" (TotalLogisticControl, 2010), 5) "Problems of Transportation" (Edupedia World, 2015b), dan 6) "Transportation Management System (TMS)" (Edupedia World, 2015c).

**3.3 Vendor**

For the second topic (vendor), the report is based on the theme of each group. Every group were asked to set their particular topic. According to their choosen topic, each group looks for the product example that is going to be discussed. After that, the product example must be elaborated based on its vendors. As example, lecturer provides an example of product "Coat". The materials of coat product (Nachum, 2015) consist of : 1) Imitation fur, ringed the hood, imported from Thailand, 2) Cotton, made the liner, imported from China and Japan, 3) Stainless steel zippers imported from Japan, 4) Snaps imported from Germany, and 5) Shell and Fleece are imported from Taiwan.

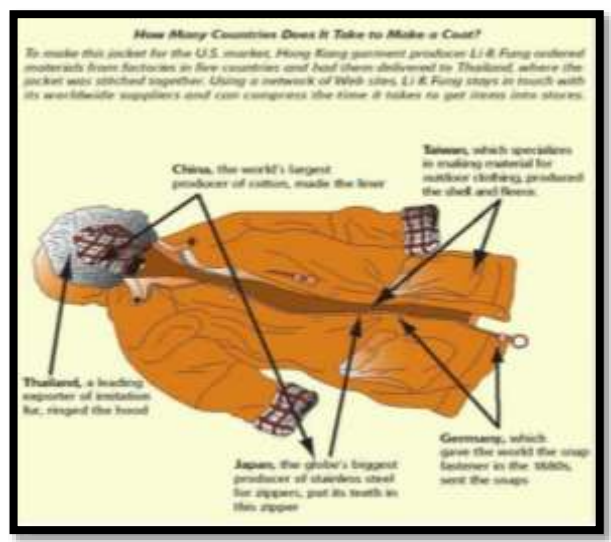

Figure 4: Product Vendor.

Based on an example from figure 1, students are asked to browse their main product. After they have their product, they need to elaborate their product. Each part of their product must be browsed to find the information about its vendor, what city and the country (if possible). Each group will add some gained information to their product.

Table 1: The number of new references.

| Group →<br>Class | A | B | C | D | E | F | Total |
|---|---|---|---|---|---|---|---|
| I | 1 | 2 | 3 | 2 | 4 | 1 | 13 |
| II | - | 1 | 3 | 5 | 5 | - | 14 |
| III | 3 | 0 | 6 | 4 | - | - | 13 |
| IV | 5 | 5 | 1 | 1 | 2 | 1 | 15 |
| | | | | | | | 55 |

Every class has 4 (four) to 6 (six) groups. Each group able to collect 1 (one) to 6 (six) references. Every class collect new references from 13 to 15. For all students there are 55 course materials in total. Among those 55 references, lecture need to select the most representative course material for all students.

### 3.4 Procurement

The third topic is "procurement", this topic videos such as: 1) Logistics - Procurement Key steps of the Purchasing Process (Supplychainlog, 2013a), 2) Logistics Procurement Process (Supplychainlog, 2013b), 3) Logistics Procurement Sourcing processes Supplier selection (Supplychainlog, 2013d), 4) Logistics Procurement processes (Supplychainlog, 2013c), 5) Supply Chain, Procurement & Logistics Consultants (BLTrecruitment, 2012), dan 6) Logistics Procurement Process by Purchasing Insight (Supplychainlog, 2013c). The example of procurement course material could be seen in Figure 5.

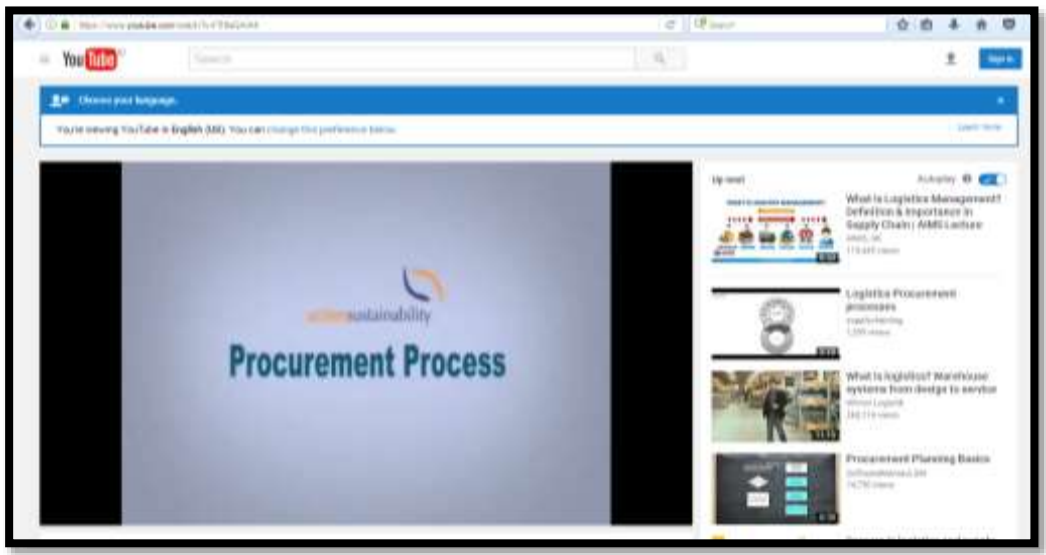

Figure 5: Procurement Course Materals Examples.

When students open the example of course materials in YouTube, they able to see related videos in right panel. For example of "procurement process" could be seen in figure 5.

## 4. Conclusion and Future Recommendation

Based on the results and discussions that have been authors described in the sections above, it can be taken a number of conclusions, as follows: 1) Social media, YouTube, is a very powerful medium in helping both lecturers and students in exploring lecture materials. In this research applied to the subject of "business modeling - supply chain management", and 2) The useness of social media facebook that is combined with YouTube has transformed the rigid learning style of the class into a wider and globalized learning environment.

## Acknowledgement

The publication of this paper is under scholarship of the "Universitas Bna Darma".